# A unified formula for the field synergy principle

**Yalin Cui**

School of Energy Science and Engineering, Harbin Institute of Technology, Harbin 150001, China
*Corresponding author: 17B302004@stu.hit.edu.cn

**Abstract**

The field synergy principle has three criteria to qualitatively describe the essence of convective heat transfer enhancement. However, in practice these criteria are difficult to be applied to convective heat transfer analysis, because there are no corresponding indicators available to quantitatively describe them. Based on these three criteria, a unified formula for the field synergy principle was developed in this study using probabilistic techniques, which is applicable to incompressible flows with constant properties in both laminar and turbulent flow regimes. The formula contains three categories of non-dimensional indicators corresponding to the three criteria of the field synergy principle respectively, including domain-averaged cosine of synergy angle, the Pearson linear correlation coefficients between the scalar functions contained in the energy governing equation of convective heat transfer, and the variation coefficients of these functions. The physical meanings of these indicators for the field synergy principle and their connections with the known heat transfer enhancing mechanisms were then discussed. Based on the unified formula, an improved analytical system for the field synergy principle was proposed, which allows an efficient and quantitative analysis of all single-phase constant-property convective heat transfer phenomena. This new system overcomes the limitation of the conventional field synergy analytical system that mainly analyzes the convective heat transfer mechanism from the perspective of synergy angle.

## 1. Introduction

The objective of heat transfer enhancing technology is to increase heat transfer rate under the constraints of pumping power and/or heat exchanging space [1-2]. Since the 1970s, increasing concerns and economic investments have been devoted to this area [3-5], and this trend is continuing. The mechanism of convective transfer enhancement is traditionally explained qualitatively from different aspects [6-8], such as mixing fluid between the wall and core region, reducing thermal boundary layer thickness, raising turbulent flow intensity, etc. An analytical tool that could provide unified and quantitative explanations of all convective heat transfer phenomena is crucial to promote the development of convective heat transfer enhancement technology. To achieve this goal, a unified theory was proposed [9] to reveal the essence of heat transfer enhancement, which is called field synergy principle (FSP). The FSP was later detailed specifically as three general criteria [10], and extended from parabolic flow to elliptic flow [11], and from incompressible flow to compressible flow [12]. Generally, the most-frequently-used FSP analytical system (referred to as the conventional FSP hereafter) mainly uses the synergy angle as a primary indicator of synergy degree to analyze the mechanism of convective heat transfer enhancements, and its main viewpoint is that a smaller synergy angle could result in a larger Nusselt number (**Nu**), when the Reynolds number (**Re**) and Prandtl number (**Pr**) are constant. Many literatures on the applications of the conventional FSP analytical system have been published in the recent decades. For example, Tao et al. [13] observed that the single-phase convective heat transfer enhancement in the laminar flow regime could be explained from the perspective of the FSP. He et al. [14] conducted parameter studies on the heat transfer characteristics of finned tube banks using a laminar flow model. The design rule for slotted fins, "front coarse, rear dense", was explained as the improved synergy in the rear part of the fin. Shen and Liu [15] studied the convective heat transfer characteristics of unsaturated porous media with the guidance of the FSP, and they found that a smaller synergy angle produces stronger heat transfer when the product of the magnitudes of the temperature gradient and velocity vector is constant. Cheng et al. [16] applied the FSP to the heat transfer analysis of a three-dimensional (3-D) rectangular channel with flush-mounted heat sources in laminar flow regime. Chen et al. [17] introduced three field synergy numbers for heat, mass and momentum transfer respectively. The conventional FSP was also successfully applied in mechanism analysis of lithium ion battery [18], longitudinal vortices [19-22], tube with twisted tape [23] and heat exchanger [24-25] etc.

The aforementioned literature survey shows that the conventional FSP were well applied to convective heat transfer analysis [26]. However, the conventional FSP was reported to be not applicable in some cases. For example, in the laminar flow regime, Tao et al. [27] carried out numerical studies on the heat transfer characteristics of wavy fin heat exchangers. They observed that the domain-averaged **Nu** does not always increase as expected with decreasing domain-averaged synergy angle. Guo et al. [28] observed that the synergy number **FC** is more appropriate to represent the synergy degree than the synergy angle. In the turbulent flow regime, the conventional FSP analytical system performs rather unsatisfactorily. Habchi et al. [29] studied three duct flow configurations using the FSP, and they observed that the local synergy angle cannot characterize the local **Nu**. Zhu et al. [30] studied the relation between the local-average-weight synergy angle and the local **Nu** in two-parallel plates with irregular boundary conditions, and it was observed that the domain-averaged synergy angle in the cross section cannot reflect the local **Nu**.

The limitations of the conventional FSP analytical system can be summarized as the following three aspects, which hinder the broad applications of this theory to convective heat transfer analysis. (1) Except the criterion relative to synergy angle, the other two criteria are neglected, which makes the analytical result one-sided. (2) Synergy angle cannot quantitatively reflect the synergy degree [21], and it even fails to predict the variational trend of synergy degree in some cases [26, 28-29], especially in turbulent flow regime. (3) Although **Fc** can represent the overall synergy degree of a thermal-flow field, it is difficult to use one indicator for in-depth heat transfer analysis.

To overcome these limitations, a unified formula for the FSP was developed in this study to incorporate all the three criteria into the analytical system, whereby improving the performance of the FSP in convective heat transfer analysis.

## 2. FSP and its three criteria

### 2.1. Brief review of the FSP

Integrating the energy equation of a two-dimensional (2-D) flat-plate laminar flow over the thermal boundary layer leads to the following equation [9]:

$$\mathbf{Nu} = \mathbf{RePr}\int_0^1 u\left|\nabla(T)\right|\cos(\theta)\mathrm{d}\tilde{l} \quad (1)$$

where $\tilde{u}$, $|\nabla(\tilde{T})|$, $\theta$ and $\tilde{l}$ are the non-dimensional velocity magnitude, non-dimensional temperature gradient, synergy angle, and non-dimensional distance perpendicular to the wall, respectively.

Extending the integral domain of Eq. (1) from the thermal boundary layer to the whole fluid domain [11], Eq. (2) is obtained:

$$\mathbf{Nu} = \mathbf{RePr}\iint_D u\left|\nabla(T)\right|\cos(\theta)\mathrm{d}\tilde{A} \tag{2}$$

where $\tilde{A}$ is non-dimensional area.

From Eq. (2), it can be concluded that **Nu** is determined by **Re**, **Pr** and the integral term $\iint_D \tilde{u}\left|\nabla(\tilde{T})\right|\cos(\theta)\,\mathrm{d}\tilde{A}$ . Tao et al. [13] noted that the term **RePr** is equal to **Nu** under fully synergy conditions (the isotherms are always perpendicular to the velocity vectors). However, **Re** and **Pr** are always constrained by working conditions, such as the given mass flow rate, size of heat exchangers, pumping power and working fluid; thus, their potential for heat transfer enhancement is limited. An efficient way of heat transfer enhancement is increasing $\iint_D \tilde{u}\left|\nabla(\tilde{T})\right|\cos(\theta)\,\mathrm{d}\tilde{A}$, i.e., the synergy number **Fc** (Eq. (3)). **Fc** is equal to unity under fully synergy condition and it is much smaller than unity for most convective heat transfer problems [10], implying that there is a large room open for convective heat transfer enhancement.

$$\mathbf{Fc}=\frac{\mathbf{Nu}}{\mathbf{RePr}}=\iint_D u\left|\nabla(T)\right|\cos(\theta)\mathrm{d}\tilde{A}. \tag{3}$$

### 2.2. Three criteria of the FSP

Three general criteria were proposed to qualitatively indicate how to increase **Fc** [11], and they provide a general insight into convective heat transfer enhancement from the perspective of field synergy. They include (1) the first criterion, "the intersection angle between the velocity and the temperature gradient/heat flow should be as small as possible", (2) the second criterion, "the local values of the three scalar fields should all be simultaneously large", (3) the third criterion, "the velocity and temperature profiles at each cross section should be as uniform as possible". The first and third criteria describe the fields' own characteristics, while the second criterion describes the relations between the three scalar fields. Therefore, these three criteria are parallel with each other and should be simultaneously satisfied as far as possible to enhance convective heat transfer.

To achieve a substantial improvement of the performance of FSP in convective heat transfer analysis, it is imperative to quantize the three criteria with different indicators and unify these indicators into the formula for **Nu**. Based on this idea, a unified formula for the FSP is developed in this study.

## 3. Unified formula for the FSP

In this section, a non-dimensional indicator named synergy coefficient (**SC**) is defined to describe the synergy degree between different scalar fields, and the calculation of **SC** by means of probabilistic method is subsequently provided. Then, a unified formula for the FSP is derived based on the three criteria for the FSP.

### 3.1. Synergy coefficient

To facilitate the formula derivation and result discussion, **SC** is defined in Eq. (4), which represents the synergy degree between different scalar function fields (such as $|\nabla(T)|$, $u$, $\cos(\theta)$ of a flow field) in terms of achieving a high integral value of their product over a specific flow domain. It is readily to deduce from Eq. (4) that if each integral value of $B_1$, $B_2$, $B_3$ … $B_n$ over Ω is constant, **SC** ($B_1$, $B_2$ … $B_n$) should be as large as possible to get a larger $\iiint_\Omega B_1B_2\cdots B_n\mathrm{d}V$ .

$$\mathbf{SC}(B_1,\ B_2\ \dots\ B_n)=\frac{\iiint_\Omega B_1B_2\cdots B_n\mathrm{d}V\Big/V_\Omega}{(\iiint_\Omega B_1\mathrm{d}V\Big/V_\Omega)(\iiint_\Omega B_2\mathrm{d}V\Big/V_\Omega)\cdots(\iiint_\Omega B_n\mathrm{d}V\Big/V_\Omega)} \tag{4}$$

where $B_1$, $B_2$, $B_3$ … and $B_n$ are different scalar functions defined in the 3-D flow domain Ω with volume $V_\Omega$.

If we divide the 3-D domain Ω into n sub-regions and take the mean values of function $B$, function $C$ and an extra function $BC$ (the product of $B$ and $C$) in each sub-region as samples of function $B$, function $C$ and the extra function $BC$ in Ω, respectively. Then three sample spaces, $S$ ($B$), $S$ ($C$) and $S$ ($BC$), can be obtained in Ω. Assume that the probability of the sample in each sub-region is equal to the ratio of the volume of the sub-region to $V_\Omega$, and if n is large enough, it is easy to get the expected values and the variances of $S$ ($B$), $S$ ($C$) and $S$ ($BC$) in Ω. Therefore, we have the following:

$$\mathbf{SC}(B,C)=\frac{\iiint_\Omega BC\mathrm{d}V\Big/V_\Omega}{\left(\iiint_\Omega B\mathrm{d}V\Big/V_\Omega\right)\left(\iiint_\Omega B\mathrm{d}V\Big/V_\Omega\right)}=\frac{E(BC)}{E(B)E(C)}. \tag{5}$$

According to probabilistic knowledge, it readily to get the following:

$$\mathbf{SC}(B,C)=1+\mathbf{r}(B,C)\mathbf{C}\cdot\mathbf{V}(B)\mathbf{C}\cdot\mathbf{V}(C) \tag{6}$$

where **C·V** is the variation coefficient that reflects the dispersion degree of the sample distribution in the sample space. **r** ($B$, $C$) is the Pearson correlation coefficient, which reflects the linear correlation degree between $S$ ($B$) and $S$ ($C$).

When $m$ is a constant, it is easy to deduce the following according to Eq. (5).

$$\mathbf{SC}(B,mC)=\mathbf{SC}(B,C). \tag{7}$$

### 3.2. Derivation of the unified formula

In this section, the unified formula for the FSP is derived in the 3-D fluid domain Ω. As illustrated in Fig. 1, Ω is surrounded by three types of boundaries that commonly exist in practice, including the wall boundaries ($S_w$) where convective heat transfer occurs, the other wall boundaries where convective heat transfer can be neglected, and the fluid boundaries where heat conduction in the boundary normal direction (the flow direction) can be neglected (as long as the Péclet number is greater than 100 [11]).

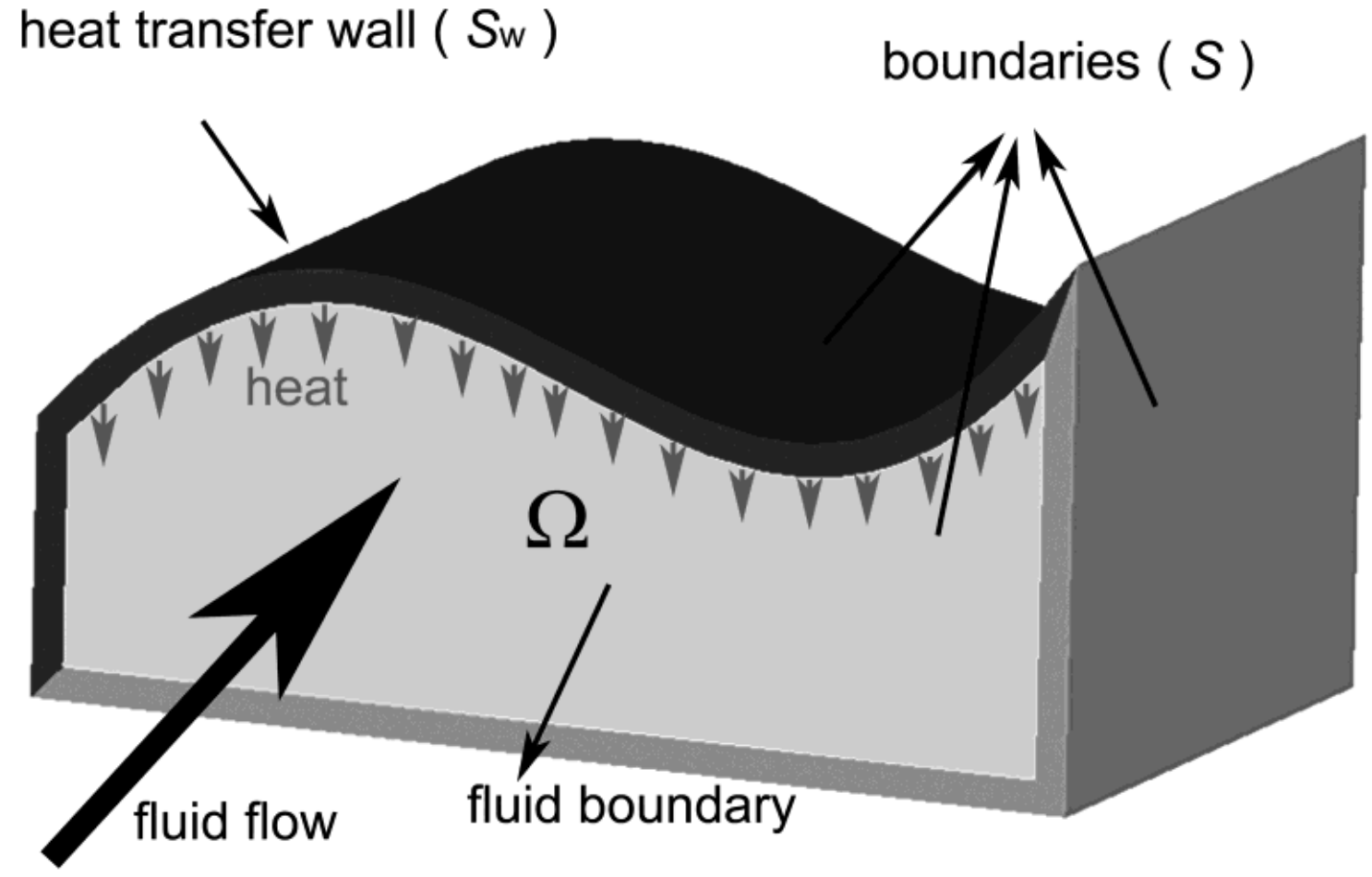

**Fig. 1. Schematic for convective heat transfer between walls and fluid.**

For most convective heat transfer problems encountered in practice, it is assumed: (a) steady state, (b) incompressible flows with constant fluid properties, (c) negligible viscous dissipation and flow work and (d) no internal heat sources. The scalar-form energy equation is as follows:

$$\rho u C_p \cos(\theta)\left|\nabla(T)\right| = \frac{\partial}{\partial x}\left(\lambda_{\mathrm{eff}}\frac{\partial T}{\partial x}\right)+\frac{\partial}{\partial y}\left(\lambda_{\mathrm{eff}}\frac{\partial T}{\partial y}\right)+\frac{\partial}{\partial z}\left(\lambda_{\mathrm{eff}}\frac{\partial T}{\partial z}\right). \tag{8}$$

Integrating the energy equation over Ω, we have the following:

$$\iiint_{\Omega}\rho u C_p \cos(\theta)\left|\nabla(T)\right|\mathrm{d}V = \iiint_{\Omega}\frac{\partial}{\partial x}\left(\lambda_{\mathrm{eff}}\frac{\partial T}{\partial x}\right)+\frac{\partial}{\partial y}\left(\lambda_{\mathrm{eff}}\frac{\partial T}{\partial y}\right)+\frac{\partial}{\partial z}\left(\lambda_{\mathrm{eff}}\frac{\partial T}{\partial z}\right)\mathrm{d}V. \tag{9}$$

Using Gauss's law to reduce the integration dimension yields the following form:

$$\iiint_{\Omega}\rho u C_p \cos(\theta)\left|\nabla(T)\right|\mathrm{d}V = \oiint_{S}\lambda_{\mathrm{eff}}\frac{\partial T}{\partial x}\mathrm{d}y\mathrm{d}z+\lambda_{\mathrm{eff}}\frac{\partial T}{\partial y}\mathrm{d}x\mathrm{d}z+\lambda_{\mathrm{eff}}\frac{\partial T}{\partial z}\mathrm{d}x\mathrm{d}y \tag{10}$$

where $S$ represents the bounding surface of the fluid domain.

Considering that both the turbulent heat flux within the fluid layer adjacent to the wall and the heat conduction in the fluid boundaries along the normal directions can be neglected, we have the following:

$$Q_{\mathrm{w}} = \iint_{S_w} q_{\mathrm{n}}\mathrm{d}A = \oiint_{S}\lambda_{\mathrm{eff}}\frac{\partial T}{\partial x}\mathrm{d}y\mathrm{d}z+\lambda_{\mathrm{eff}}\frac{\partial T}{\partial y}\mathrm{d}x\mathrm{d}z+\lambda_{\mathrm{eff}}\frac{\partial T}{\partial z}\mathrm{d}x\mathrm{d}y = \iiint_{\Omega}\rho u C_p \cos(\theta)\left|\nabla(T)\right|\mathrm{d}V \tag{11}$$

where $Q_{\mathrm{w}}$ represents the total heat transfer rate between the fluid and the heat transfer walls $S_{\mathrm{w}}$, and $q_{\mathrm{n}}$ represents the wall heat flux perpendicular to wall $S_{\mathrm{w}}$.

Then, we have the following equation for the convective heat transfer coefficient $h_{\mathrm{w}}$:

$$h_{\mathrm{w}} = \frac{Q_{\mathrm{w}}}{(\overline{T}_{\mathrm{w}} - T_{\mathrm{b}})A_{\mathrm{w}}} = \frac{\iiint_{\Omega}\rho u C_p \cos(\theta)\left|\nabla(T)\right|\mathrm{d}V}{(\overline{T}_{\mathrm{w}} - T_{\mathrm{b}})A_{\mathrm{w}}} \tag{12}$$

where $\overline{T}_{\mathrm{w}} = \iint_{S_{\mathrm{w}}} T_{\mathrm{w}}dA \big/ A_{\mathrm{w}}$ , $T_{\mathrm{b}} = \iiint_{\Omega} uT\mathrm{d}V \big/ \iiint_{\Omega} u\mathrm{d}V$ , $A_{\mathrm{w}}$ is the area of $S_{\mathrm{w}}$ .

Introduce the characteristic length $L'$, we obtain the following equation for **Nu**:

$$\mathbf{Nu} = \frac{h_{\mathrm{w}}L'}{\lambda} = \frac{L'}{\lambda(\overline{T}_{\mathrm{w}} - T_{\mathrm{b}})A_{\mathrm{w}}}\iiint_{\Omega}\rho u C_p \cos(\theta)\left|\nabla(T)\right|\mathrm{d}V = \mathbf{RePr}\frac{\iiint_{\Omega}\rho u C_p \cos(\theta)\left|\nabla(T)\right|\mathrm{d}V}{\left(\iiint_{\Omega}\rho u C_p\mathrm{d}V \big/ V_{\Omega}\right)\left(\overline{T}_{\mathrm{w}} - T_{\mathrm{b}}\right)A_{\mathrm{w}}} \tag{13}$$

where: $\mathbf{Re}=\dfrac{u_{\infty}L'}{\upsilon}$ , $\mathbf{Pr}=\dfrac{\upsilon\rho C_p}{\lambda}$ , $u_{\infty}=\dfrac{\iiint_{\Omega}u\mathrm{d}V}{V_{\Omega}}$ , $\lambda$ is thermal conductivity, $\upsilon$ is kinematic viscosity, $\rho$ is density, $C_p$ is constant-pressure specific heat.

Eq. (13) can be deformed to the following form:

$$\mathbf{Nu}=\mathbf{RePr}\frac{\iiint_{\Omega}\left|\nabla(T)\right|\mathrm{d}V}{\left(\overline{T}_{\mathrm{w}}-T_{\mathrm{b}}\right)A_{\mathrm{w}}}\frac{\iiint_{\Omega}\rho uC_p\cos(\theta)\left|\nabla(T)\right|\mathrm{d}V\big/V_{\Omega}}{\left(\iiint_{\Omega}\rho uC_p\cos(\theta)\mathrm{d}V\big/V_{\Omega}\right)\left(\iiint_{\Omega}\left|\nabla(T)\right|\mathrm{d}V\big/V_{\Omega}\right)}\frac{\iiint_{\Omega}\rho uC_p\cos(\theta)\mathrm{d}V\big/V_{\Omega}}{\left(\iiint_{\Omega}\rho uC_p\mathrm{d}V\big/V_{\Omega}\right)\left(\iiint_{\Omega}\cos(\theta)\mathrm{d}V\big/V_{\Omega}\right)}$$
$$\times\left(\iiint_{\Omega}\cos(\theta)\mathrm{d}V\big/V_{\Omega}\right). \tag{14}$$

Then, with the help of **SC** (Eq. (6)), three key non-dimensional parameters are defined as follows:

$$\boldsymbol{\varphi}_{\mathrm{HA}}=\frac{\iiint_{\Omega}\rho uC_p\cos(\theta)\mathrm{d}V\big/V_{\Omega}}{\left(\iiint_{\Omega}\rho uC_p\mathrm{d}V\big/V_{\Omega}\right)\left(\iiint_{\Omega}\cos(\theta)\mathrm{d}V\big/V_{\Omega}\right)}\left(\iiint_{\Omega}\cos(\theta)\mathrm{d}V\big/V_{\Omega}\right)=\mathbf{SC}(\rho uC_p,\cos(\theta))\overline{\cos(\theta)}$$
$$=(1+\mathbf{r}(\rho uC_p,\cos(\theta))\mathbf{C}\cdot\mathbf{V}(\rho uC_p)\mathbf{C}\cdot\mathbf{V}(\cos(\theta)))\overline{\cos(\theta)} \tag{15}$$

$$\boldsymbol{\varphi}_{\mathrm{HT}}=\frac{\iiint_{\Omega}\rho uC_p\cos(\theta)\left|\nabla(T)\right|\mathrm{d}V\big/V_{\Omega}}{\left(\iiint_{\Omega}\rho uC_p\cos(\theta)\mathrm{d}V\big/V_{\Omega}\right)\left(\iiint_{\Omega}\left|\nabla(T)\right|\mathrm{d}V\big/V_{\Omega}\right)}=\mathbf{SC}(\rho uC_p\cos(\theta),\left|\nabla(T)\right|)=1+\mathbf{r}(\rho uC_p\cos(\theta),\left|\nabla(T)\right|)$$
$$\times\mathbf{C}\cdot\mathbf{V}(\rho uC_p\cos(\theta))\mathbf{C}\cdot\mathbf{V}(\left|\nabla(T)\right|) \tag{16}$$

$$\mathbf{k}'=\frac{\iiint_{\Omega}\left|\nabla(T)\right|\mathrm{d}V\big/A_{\mathrm{w}}}{\left(\overline{T}_{\mathrm{w}}-T_{\mathrm{b}}\right)} \tag{17}$$

where $\overline{\cos(\theta)}$ is the domain-averaged cosine of synergy angle; $\boldsymbol{\varphi}_{\mathrm{HA}}$ and $\boldsymbol{\varphi}_{\mathrm{HT}}$ are two parameters to facilitate the formula derivation; $\mathbf{k}'$ is the correction factor required to convert the domain-averaged temperature difference $\iiint_{\Omega}\left|\nabla(T)\right|\mathrm{d}V\big/A_{\mathrm{w}}$ to $\left(\overline{T}_{\mathrm{w}}-T_{\mathrm{b}}\right)$, so that the equation can coordinate with **Nu** which is defined by $\left(\overline{T}_{\mathrm{w}}-T_{\mathrm{b}}\right)$. Because $\iiint_{\Omega}\left|\nabla(T)\right|\mathrm{d}V\big/A_{\mathrm{w}}$ and $\left(\overline{T}_{\mathrm{w}}-T_{\mathrm{b}}\right)$ both calculate the average temperature difference between wall $S_{\mathrm{w}}$ and fluid in $\Omega$, the value of $\mathbf{k}'$ is quite close to unity.

Since the fluid properties are assumed to be constant, according to Eq. (8), we have:

$$\boldsymbol{\varphi}_{\mathrm{HA}}=(1+\mathbf{r}(u,\cos(\theta))\mathbf{C}\cdot\mathbf{V}(u)\mathbf{C}\cdot\mathbf{V}(\cos(\theta)))\overline{\cos(\theta)} \tag{18}$$

$$\boldsymbol{\varphi}_{\mathrm{HT}}=1+\mathbf{r}(u\cos(\theta),\left|\nabla(T)\right|)\mathbf{C}\cdot\mathbf{V}(u\cos(\theta))\mathbf{C}\cdot\mathbf{V}(\left|\nabla(T)\right|) \tag{19}$$

Substituting Eqs. (15), (16) and (17) into Eq. (14), we get the following:

$$\mathbf{Nu}=\mathbf{RePrk}'\boldsymbol{\varphi}_{\mathrm{HA}}\boldsymbol{\varphi}_{\mathrm{HT}}. \tag{20}$$

Substituting Eqs. (18) and (19) into Eq. (20), we get the unified formula for the FSP:

$$\mathbf{Nu}=\mathbf{RePrk}'\overline{\cos(\theta)}\left(1+\mathbf{r}(u,\cos(\theta))\mathbf{C}\cdot\mathbf{V}(u)\mathbf{C}\cdot\mathbf{V}(\cos(\theta))\right)\left(1+\mathbf{r}(u\cos(\theta),\left|\nabla(T)\right|)\mathbf{C}\cdot\mathbf{V}(u\cos(\theta))\mathbf{C}\cdot\mathbf{V}(\left|\nabla(T)\right|)\right). \tag{21}$$

## 4. In-depth interpretation of the unified formula

In this section, we will discuss the physical meanings of the indicators involved in the unified formula for the FSP, and then introduce an improved FSP analytical system.

According to Eq. (3) and Eq. (21), we have the following:

$$\mathbf{Fc}=\mathbf{k}'\overline{\cos(\theta)}\left(1+\mathbf{r}(u,\cos(\theta))\mathbf{C}\cdot\mathbf{V}(u)\mathbf{C}\cdot\mathbf{V}(\cos(\theta))\right)\left(1+\mathbf{r}(u\cos(\theta),\left|\nabla(T)\right|)\mathbf{C}\cdot\mathbf{V}(u\cos(\theta))\mathbf{C}\cdot\mathbf{V}(\left|\nabla(T)\right|)\right). \quad (22)$$

The product of **Re** and **Pr** in Eq. (21) can be regarded as the potential of convective heat transfer, i.e. the maximum **Nu** virtually achievable [13]. The value of **k′** is close to unity, hence **k′** has little influence on **Nu**. Thus, according to Eq. (22), **Fc** is determined by three categories of non-dimensional indicators including the first category: $\overline{\cos(\theta)}$; the second category: **r** ($u$, cos($\theta$)) and **r** ($u$cos($\theta$), $|\nabla(T)|$); the third category: **C·V** ($u$), **C·V** (cos($\theta$)), **C·V** ($u$cos($\theta$)) and **C·V** ($|\nabla(T)|$). These FSP indicators correspond to the three criteria of the FSP, respectively, and their physical explanations are presented in the following.

### 4.1. Indicator corresponding to the first criterion of the FSP

$\overline{\cos(\theta)}$ is the indicator corresponds to the first criterion of the FSP, and it may be explained as follows: as shown in Fig. 2 (a), $\overline{\cos(\theta)}$ is quite close to zero in most cases, owning to the fact that velocity vector and temperature gradient vector are approximately parallel to and perpendicular to heat transfer wall, respectively. However, $\overline{\cos(\theta)}$ could be increased or decreased greatly when the flow swirls or fluctuates because of the interactions with heat transfer enhancement components, accompanying the transporting of fluid from the wall into the free stream and vice versa, as shown in Fig. 2 (b). Therefore, increasing $\overline{\cos(\theta)}$ (decreasing synergy angle) generally accompanies the enhancement of secondary flow, which could result in convective heat transfer argumentation introduced by fluid exchange between the wall and core region. However, enhancing fluid mixing between the wall and free stream to argument convective heat transfer is generally more effective in laminar flow regime as compared with in the turbulent flow regime. This is because in the laminar flow regime, heat will be transferred within fluid by pure conduction and the heat transfer capacity will be at the lowest level if there is no fluid exchange by convection in the temperature gradient direction. As for the case in the turbulent flow regime, reducing the thermal resistance posed by the boundary layer to enhance convective heat transfer is generally considered more effective as compared with fluid mixing, because heat could be transferred effectively by local instantaneous fluid pulsation (the effective conductivity is much larger than molecular thermal conductivity) in the region away from boundary layer. Therefore, synergy angle is capable of evaluating the synergy degree (i.e. **Nu** when **Re** and **Pr** are constant) of convective heat transfer in the laminar flow regime. This explains why the synergy-angle-based conventional FSP analytical system performs much better in the laminar flow regime than in the turbulent flow regime, which has been well presented in the literature survey of the introduction section.

### 4.2. Indicators corresponding to the second criterion of the FSP

These indicators include **r** ($u$, cos($\theta$)) and **r** ($u$cos($\theta$), $|\nabla(T)|$). Pearson correlation coefficient is to describe the linear degree between two different data sets (groups of samples). Thus, **r** ($u$, cos($\theta$)) and **r** ($u$cos($\theta$), $|\nabla(T)|$) can be taken as indicators for the second criterion of FSP. This criterion could be explained as follows: generally, the fluid velocity field is much stronger in the region away from the heat transfer walls due to the existence of boundary layer, whereas the situation is reversed for the temperature gradient field. As for synergy angle field, the value of cos($\theta$) gets close to zero in the near wall regions as the temperature vector gets more perpendicular to velocity vector. Thereby, **r** ($u$cos($\theta$), $|\nabla(T)|$) is negative in most cases. When flow impinges heat transfer wall with small angle (angle between velocity vector and wall normal direction), the local $u$, cos($\theta$) and $|\nabla(T)|$ fields will be simultaneously strong in this fluid-impinging-wall region (Fig. 2 (c)), and as a result **r** ($u$cos($\theta$), $|\nabla(T)|$) will be increased along with the local reduction of thermal boundary layer thickness and the resulting convective heat transfer enhancement. Therefore, **r** ($u$cos($\theta$), $|\nabla(T)|$) is closely related to the interaction between fluid and heat transfer wall, and it can be used to evaluate the convective heat transfer enhancing technologies that are achieved by increasing the interaction between wall and fluid, such as jet impingement, longitudinal vortex generator, corrugated tube, louvered fins, etc.

As for **r** ($u$, cos ($\theta$)), it is to a great extent determined by synergy angle field as the velocity magnitude distribution along the wall normal direction is relatively fixed. When the local cos ($\theta$) becomes larger in the near wall regions, **r** ($u$, cos ($\theta$)) is decreased, while **r** ($u$cos($\theta$), $|\nabla(T)|$) is increased. Therefore, **r** ($u$, cos ($\theta$)) generally has a negative correlation with **r** ($u$cos($\theta$), $|\nabla(T)|$).

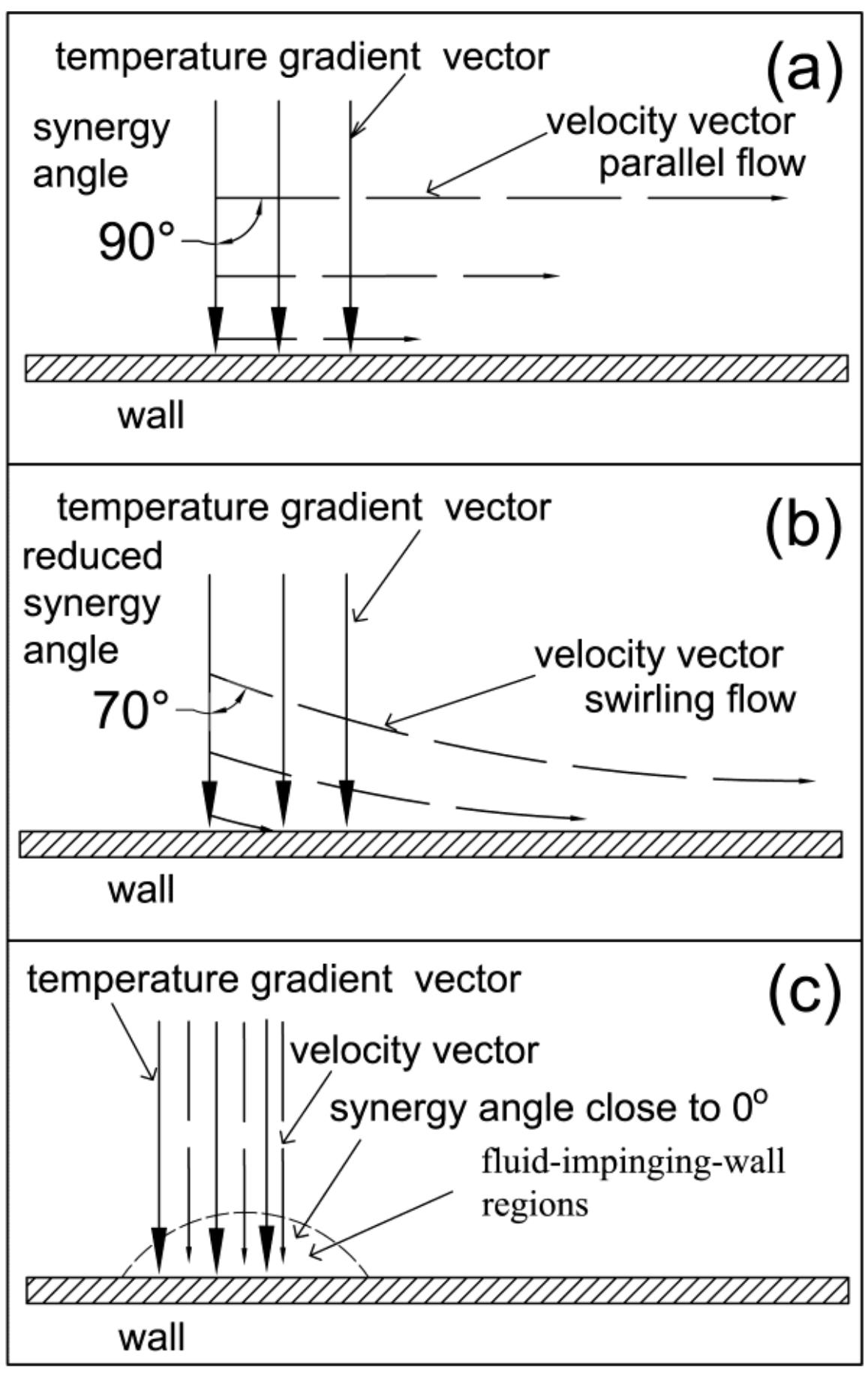

**Fig.2. Parallel flow, swirling flow and jet impingement.**

### 4.3. Indicators corresponding to the third criterion of the FS

This category of FSP indicator includes **C·V** ($u$), **C·V** (cos($\theta$)), **C·V** ($u$cos($\theta$)) and **C·V** ($|\nabla(T)|$). Variation coefficient describes the dispersion degree of a data set, and a small dispersion means a good uniformity. Therefore, they can be treated as indicators for the third criterion. Because **r** ($u\cos(\theta)$, $|\nabla(T)|$) is negative in most cases, according to Eq. (22), **C·V** ($u\cos(\theta)$) and **C·V** ($|\nabla(T)|$) should be as small as possible to enhance the heat transfer, i.e., the fields of $u\cos(\theta)$ and $|\nabla(T)|$ should be as uniform as possible. It was reported in [10] that improving the uniformity of the temperature field could lead to convective heat transfer enhancement. Generally, for a specific convective heat transfer process with fixed boundary conditions, a better uniformity of temperature gradient field (small value of **C·V** ($|\nabla(T)|$)) means that heat is more effectively transferred from the wall to the regions away from the walls. Therefore, the technologies based on increasing the effective conductivity of fluid, e.g. promoting turbulent intensity, adding high-thermal conductivity particles into fluid, liquid metal convection heat transfer, etc. are closely related to **C·V** ($|\nabla(T)|$). As for **C·V** ($u\cos(\theta)$), it decreases with the increase of $\cos(\theta)$ in the near wall regions where the local velocity field is weak. Convective heat transfer enhancement in the hydraulic and/or thermal entry region may be related to **C·V** ($u$) and **C·V** ($|\nabla(T)|$), considering that **C·V** ($u$) and **C·V** ($|\nabla(T)|$) are very small in this region. Overall, the importance of this group of indicators for convective heat transfer enhancement need to be further explored through much more case studies.

### 4.4. Improved FSP analytical system

Consequently, an improved FSP analytical system to analyze the convective heat transfer mechanism is established with its analytical process schematically shown in Fig. 3. It provides unified and quantitative explanations for all single-phase constant-property convective heat transfer phenomena from the perspective of field synergy. Generally, for a specific convective heat transfer problem, only several of the FSP indicators have prominent effects on convective heat transfer behaviors based on their contributions to the variation of **Nu**, which could help us simplify the FSP analysis. By judging which group the most significant FSP indicator for a convective heat transfer process belongs to, this system classifies convective heat transfer enhancement technologies into three modes, including decreasing synergy angle, increasing the linear correlation coefficient between $u$, $\cos(\theta)$ and $|\nabla(T)|$ and improving the field uniformities of $u$, $\cos(\theta)$ and $|\nabla(T)|$, leading to a unified understanding of all single-phase convective heat transfer phenomena. On the other hand, because this system is based on statistical indicators, its advantages in heat transfer mechanism analysis over the conventional methods would be more prominent in complex heat-flow-conjugated problems, in which the conventional methods are difficult to be applied, such as flow under complicated boundary conditions (it is difficult to make a qualitative analysis just through observing the contours of temperature, velocity or other related variables). However it should be noted that, as a new analytical system for convective heat transfer, great efforts are required to further explore and refine this theory in the following aspects: (1) although some work has been done in this study, it needs to further clarify the physical connotations of the FSP indicators through more case studies for better applications in convective heat transfer analysis, (2) explore the possible connections of the FSP indicators with the indicator for thermal-hydraulic performance and the way how to apply the analytical results to optimize convective heat transfer.

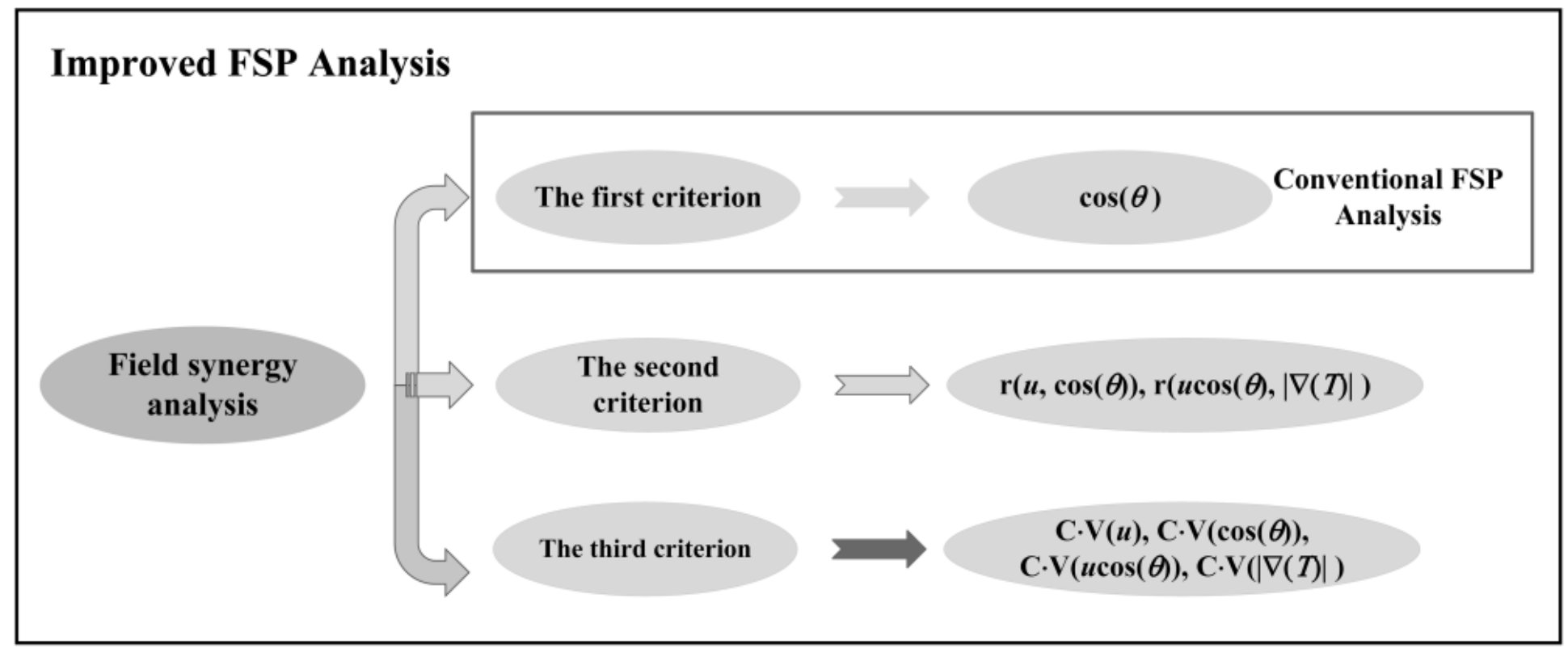

**Fig. 3. The improved FSP analytical system.**

## 5. Conclusions

A unified formula was derived based on the three general criteria of the FSP by means of probabilistic techniques. The unified formula is applicable to incompressible flows with constant fluid properties in both laminar and turbulent flow regimes. The unified formula quantifies the three criteria of the FSP with three categories of non-dimensional indicators, including: (1) $\overline{\cos(\theta)}$ (the first criterion of the FSP); (2) **r** ($u$, cos($\theta$)) and **r** ($u$cos($\theta$), $|\nabla(T)|$) (the second criterion of the FSP); (3) **C·V** ($u$), **C·V** (cos($\theta$)), **C·V** ($u$cos($\theta$)) and **C·V** ($|\nabla(T)|$) (the third criterion of the FSP). Thus, convective heat transfer mechanism analysis can be conducted from the three aspects of the FSP quantitatively, overcoming the limitation of the conventional FSP analytical system which mainly takes the synergy angle as the primary indicator to describe the synergy degree. The improved FSP analytical system could provide unified and quantitative explanations for all single-phase constant-property convective heat transfer phenomena. As an analytical method based on statistical indicators, its merits over the conventional ones in convective heat transfer analysis would be more prominent for convective heat transfer problems under complicated boundary conditions.

**Acknowledgments**

It should be noted that all the novel ideas involved in this paper were proposed independently by Yalin Cui with his great efforts, and also the paper is written independently by Yalin Cui.

**Nomenclature**

$A$ – area, [m$^2$]

$C_p$ – specific heat capacity at constant pressure, [J kg$^{-1}$K$^{-1}$]

$h$ – convective heat transfer coefficient, [Wm$^{-2}$K$^{-1}$]

$L$ – length, [m]

$q_{\mathrm{n}}$ – heat flux perpendicular to the wall, [Wm$^{-2}$]

$Q_{\mathrm{w}}$ – heat transfer rate between wall $S_{\mathrm{w}}$ and fluid, [W]

Ω – fluid domain

**Non-dimensional parameters**

**C·V** – variation coefficient

**Fc** – synergy number, **Nu**/(**RePr**)

**k′** – correlation factor of temperature difference

**Nu** – Nusselt number, $hL/\lambda$

$S$ – the bounding surface of the fluid domain, [$m^{-2}$]

$S_w$ – wall boundaries where the convective heat transfer occurs, [$m^2$]

$T$ – temperature, [K]

$|\nabla(T)|$ – temperature gradient, [$Km^{-1}$]

$u$ – velocity magnitude, [$ms^{-1}$]

$u_\infty$ – average velocity of the fluid domain, [$ms^{-1}$]

$V$ – volume of fluid domain, [$m^{-3}$]

**Greek symbols**

$\theta$ – synergy angle, [°]

$\lambda$ – thermal conductivity, [$Wm^{-1}K^{-1}$]

$\mu$ – dynamic viscosity, [$Pas^{-1}$]

$\rho$ – density, [$kgm^{-3}$]

**Pr** – Prandtl number, $\mu C_p/\lambda$

**Re** – Reynolds number, $\rho uL/\mu$

**r** – Pearson correlation coefficient

**Rt** – ratio of the FSP indicator of TB to that of the ST

**SC** – synergy coefficient

**Subscripts**

b – bulk

eff – effective

w – wall

**Abbreviations**

FSP – field synergy principle

**References**

[1] Kundu B., Lee K. S., A proper analytical analysis of annular step porous fins for determining maximum heat transfer, *Energy Conversion and Management*, *110* (2016), Feb., PP. 469-480

[2] Kim M. H., et al., A novel louvered fin design to enhance thermal and drainage performances during periodic frosting/defrosting conditions, *Energy Conversion and Management*, *110* (2016), Feb., PP. 494-500

[3] Zhang Y. N., et al., Characteristics of geothermal energy obtained from a deep well in the coldest provincial capital of China, *Thermal Science*, *22* (2018), S2, PP. 673-679

[4] Zhang Y. N., et al., Exergy characteristics of rice husks, *Thermal Science*, 22 (2018), S2, PP. 429-437

[5] Yang X. J., Gao F., A new technology for solving diffusion and heat equations, *Thermal Science*, 21 (2017), 1A, PP. 246-246

[6] Atangana A., Baleanu D., New Fractional Derivatives with Nonlocal and Non-Singular Kernel: Theory and Application to Heat Transfer Model, *Thermal Science*, 20 (2016), 2, PP. 763-769

[7] Ognjanovic O., et al., Numerical aerodynamic-thermal-structural analyses of missile fin configuration during supersonic flight conditions, *Thermal Science*, 21 (2016), 6B, PP. 318-318

[8] Tekelioğlu M., Empirical mapping of the convective heat transfer coefficients with local hot spots on highly conductive surfaces, *Thermal Science*, 21 (2015), 3, PP. 153-153

[9] Guo Z. Y., et al., A novel concept for convective heat transfer enhancement, *International Journal of Heat and Mass Transfer*, *41* (1998), 14, PP. 2221-2225

[10] Guo Z. Y., et al., The field synergy (coordination) principle and its applications in enhancing single phase convective heat transfer, *International Journal of Heat and Mass Transfer*, *48* (2005), 9, PP. 1797-1807

[11] Tao W. Q., et al., Field synergy principle for enhancing convective heat transfer—its extension and numerical verifications, *International Journal of Heat and Mass Transfer*, *45* (2002), 18, PP.

3849-3856

[12] Zhang B., et al., Compressible fluid flow field synergy principle and its application to drag reduction in variable-cross-section pipeline, *International Journal of Heat and Mass Transfer*, *77* (2014), Oct., PP. 1095-1101

[13] Tao W. Q., et al., A unified analysis on enhancing single phase convective heat transfer with field synergy principle, *International Journal of Heat and Mass Transfer*, *45* (2002), 24, PP. 4871-4879

[14] He Y. L., et al., Three-dimensional numerical study of heat transfer characteristics of plain plate fin-and-tube heat exchangers from view point of field synergy principle, *International Journal of Heat and Fluid Flow*, *26* (2005), 3, PP. 459-473

[15] Shen S., et al., Analysis of field synergy on natural convective heat transfer in porous media, *International Communications in Heat and Mass Transfer*, *30* (2003), 8, PP. 1081-1090

[16] Cheng Y. P., et al., Numerical analysis of mixed convection in three-dimensional rectangular channel with fush-mounted heat sources based on field synergy principle, *International Journal for numerical methods in fluids*, *52* (2006), 9, PP. 987-1003

[17] Chen Q., et al., Field Synergy Principle for Energy Conservation Analysis and Application, *Advances in Mechanical Engineering*, 2 (2010), *11*, PP. 1652-1660

[18] He H. W., et al., Field synergy analysis and optimization of the thermal behavior of lithium ion battery packs, *Energies*, *10* (2017), 1, PP. 81-91

[19] Lei Y. G., et al., Improving the thermal hydraulic performance of a circular tube by using punched delta-winglet vortex generators, *International Journal of Heat and Mass Transfer*, *111* (2017), Aug., PP. 299-311

[20] Lotfi B., et al., An investigation of the thermo-hydraulic performance of the smooth wavy fin-and-elliptical tube heat exchangers utilizing new type vortex generators, *Applied Energy*, *162* (2015), June, PP. 1282-1302

[21] Lu G. F., Zhou G. B., Numerical simulation on performances of plane and curved winglet – Pair vortex generators in a rectangular channel and field synergy analysis, *International Journal of Thermal Sciences*, *109* (2016), Nov., PP. 323-333

[22] Habchi C, et al., Entropy production and field synergy principle in turbulent vortical flows, *International Journal of Thermal Sciences*, *50* (2011), 12, PP. 2365-2376

[23] Cao Y. P., et al., Investigation on the flow noise propagation mechanism in pipelines of shell-and-tube heat exchangers based on synergy principle of flow and sound fields, *Applied Thermal Engineering*, *122* (2017), July, PP. 339-349

[24] Zhang X. Y., et al., Numerical studies on heat transfer and flow characteristics for laminar flow in a tube with multiple regularly spaced twisted tapes, *International Journal of Thermal Sciences*, *58* (2012), 2, PP. 157-167

[25] Li Z., et al., A numerical study of laminar convective heat transfer in microchannel with non-circular cross-section, *International Journal of Thermal Sciences*, *45* (2006), 14, PP. 1140-1148

[26] Cheng Y. P., et al., Numerical design of efficient slotted fin surface based on the field synergy principle, *Numerical Heat Transfer, Part A: Applications*, *45* (2004), 6, PP. 517-538

[27] Tao Y. B., et al., Three-dimensional numerical study and field synergy principle analysis of wavy fin heat exchangers with elliptic tubes, *International Journal of Heat and Fluid Flow*, *28* (2007), 6, PP. 1531-1544

[28] Guo J. F., et al., Numerical investigations of curved square channel from the viewpoint of field

synergy principle, *International Journal of Heat and Mass Transfer*, *54* (2011), 17, PP. 4148-4151
[29] Habchi C., et al., Entropy production and field synergy principle in turbulent vortical flows, *International Journal of Thermal Sciences*, *50* (2011), 12, PP. 2365-2376
[30] Zhu X. W., Zhao J.Q., Improvement in field synergy principle: More rigorous application, better results, *International Journal of Heat and Mass Transfer*, *100* (2016), Sep., PP. 347-354